# Entropy: An inherent, nonstatistical property of any system in any state

Elias P. Gyftopoulos
Massachusetts Institute of Technology
77 Massachusetts Avenue, Cambridge, Massachusetts 02139

Entropy is the distinguishing and most important concept of our efforts to understand and regularize our observations of a very large class of natural phenomena, and yet, it is one of the most contentious concepts of physics. In this article, we review two expositions of thermodynamics, one without reference to quantum theory, and the other quantum mechanical without probabilities of statistical mechanics. In the first, we show that entropy is an inherent property of any system in any state, and that its analytical expression must conform to eight criteria. In the second, we recognize that quantum thermodynamics: (i) admits quantum probabilities described either by wave functions or by nonstatistical density operators; and (ii) requires a nonlinear equation of motion that is delimited by but more general than the Schrödinger equation, and that accounts for both reversible and irreversible evolutions of the state of the system in time. Both the more general quantum probabilities, and the equation of motion have been defined, and the three laws of thermodynamics are shown to be theorems of this equation.

## I. INTRODUCTION

Ever since Clausius postulated that "the energy of the universe is constant", and "the entropy of the universe strives to attain a maximum value", practically every scientist and engineer shares the beliefs that: (i) Thermodynamics is a statistical theory, restricted to phenomena in macroscopic systems in thermodynamic equilibrium states; and (ii) Entropy – the concept that distinguishes thermodynamics from mechanics – is a statistical measure of ignorance, ultimate disorder, dispersion of energy, erasure of information, or other causes, and not an inherent property of matter like rest mass.

These beliefs stem from the conviction that the "known laws" of mechanics (classical or conventional quantum) are the ultimate laws of physics, and from the fact that statistical theories of thermodynamics yield accurate and practical numerical results about thermodynamic equilibrium states.

Notwithstanding the conviction and excellent numerical successes, the almost universal efforts to compel thermodynamics to conform to statistical and other non-physical explanations, and to restrict it only to thermodynamic equilibrium states [1-3] are puzzling in the light of many accurate, reproducible, and nonstatistical experiences, and many phenomena that cannot possibly be described in terms of thermodynamic equilibrium states.

Since the advent of thermodynamics, many academics and practitioners have questioned the clarity, unambiguity, and logical consistency of traditional expositions of the subject. Some of the questions raised are: (i) Why is thermodynamics restricted to thermodynamic equilibrium states only, given that the universally accepted and practical statements of energy conservation and entropy nondecrease are demonstrably time dependent?; (ii) Why do we restrict thermodynamics to macroscopic systems, given that even Gibbsian statistics [4, 5], and systems in states with negative temperatures [6] prove beyond a shadow of a doubt that thermodynamics is valid for any system?; (iii) How can any of the proposed statistical expressions of entropy be accepted if, as we will see later, none conforms to the requirements that must be satisfied by the entropy of thermodynamics?; and (vi) Why do so many professionals continue to believe that thermodynamic equilibrium is a state of ultimate disorder despite the fact that both experimental and theoretical evidence indicates that such a state represents ultimate order [7, 8]?

In what follows, we prove that thermodynamics is a well founded, nonstatistical general theory of physics. We present brief summaries of two novel, intimately interrelated, and revolutionary, in the sense of Kuhn [9], expositions. The first is purely thermodynamic without any probabilities, and is discussed in Section II. In this exposition, entropy is proven to be an inherent, nonstatistical property of any system (either large or small), in any state (either thermodynamic equilibrium or not thermodynamic equilibrium). The second exposition is purely quantum mechanical, i.e., the probabilities are not mixtures of quantum and statistical probabilities, and is discussed in Section III. The evolution in time of the probabilities just cited requires an equation of motion more general than either the Schrödinger or the statistical von Neumann equation. Such an equation is discussed in Section IV, and our conclusions in Section V.

## II. A NEW EXPOSITION OF THERMODYNAMICS

### II.1 Foundations

Over the past more than three decades, a small group at the Massachusetts Institute of Technology developed a nonstatistical and nonquantum mechanical exposition of the foundations and applications of thermodynamics that applies to all systems (including one particle or one spin systems) and to both thermodynamic equilibrium and not thermodynamic equilibrium states [10].

In the new exposition, we start with the mechanical concepts of space, time, and inertial mass or force, and express the first law as follows: *Any two states $A_1$ and $A_2$ of system A may always be the end states of a process that involves no other effects external to the system except the change in elevation of a weight between $z_1$ and $z_2$, that is, solely a purely mechanical effect, and $z_1 - z_2$ depends only on $A_1$ and $A_2$.* In contrast to other expositions, it is noteworthy that this statement does not involve the concepts of energy, temperature, heat, and work, all of which are defined later.

The first law implies many rigorously proven theorems. Examples are: (i) At each state of a system there must exist a function $E$, called *energy*, such that the change of its value $E_2 - E_1$ from state $A_1$ to state $A_2$ is proportional to $z_2 - z_1$; (ii) In the course of spontaneous changes of state (changes in time in an isolated system), $E$ is invariant; and (iii) In the course of interactions, $E_2 - E_1$ must be accounted for by the energy exchanged with systems interacting with $A$, that is, an energy balance must be satisfied.

Next, depending on their evolution in time, we classify states in the seven categories encountered in mechanics, that is, unsteady, steady, nonequilibrium, equilibrium, unstable equilibrium, metastable equilibrium, and stable equilibrium, and raise the question: For given values of the energy, the volume, and the amounts of constituents of a system, are there any stable equilibrium states?

In the new exposition, the answer is given by the second law which avers that (simplified version): *For each set of values of energy E, amounts of r constituents **n**, and volume V, there exists one and only one stable equilibrium state.* It is noteworthy that the concept of stable equilibrium is what in ordinary expositions is called equilibrium or thermodynamic equilibrium, and that, in contrast to all other expositions, here the second law does not involve the concepts of heat, temperature, and entropy.

The second law cannot be derived from or explained by the "known laws" of physics either directly or statistically because these laws imply that the state of lowest energy is the only stable equilibrium state, whereas the second law avers that such a state exists for each set of values $E$, $\mathbf{n}$, $V$.

Among the many rigorously proven theorems of the two laws, one is established as follows. Upon defining a reservoir in terms of concepts that have already been introduced [10], we investigate the optimum amount of energy that can be exchanged between a weight and a composite of system $A$ and reservoir $R$ – the optimum mechanical effect. We call this optimum value *generalized available energy*, denote it by $\Omega^R$, and show that it is additive, and a generalization of the motive power of fire introduced by Carnot. It is a generalization because Carnot assumed that $A$ is also a reservoir, and we do not.

For an *adiabatic process* of system $A$ only, we show that the changes of energy $E_1 - E_2$ of $A$ and of the generalized available energy $\Omega_1^R - \Omega_2^R$ of the composite of $A$ and $R$ satisfy the relations:

$$E_1 - E_2 = \Omega_1^R - \Omega_2^R \tag{1}$$

if the process is reversible, or

$$E_1 - E_2 < \Omega_1^R - \Omega_2^R \tag{2}$$

if the process is irreversible. A *process is reversible* if both the system and its environment can be restored to their respective initial states. A *process is irreversible* if the restoration just cited is impossible.

The two properties $E$ and $\Omega^R$ determine a property of $A$ only, which is called *entropy* and is denoted by $S$. For state $A_1$, $S_1$ is evaluated by means of any reservoir $R$, a reference state $A_0$, and the expression

$$S_1 = S_0 + \frac{1}{c_R}\left[(E_1 - E_0) - (\Omega_1^R - \Omega_0^R)\right] \tag{3}$$

where $c_R$ is a well defined positive constant that depends only on the reservoir. The entropy $S$ is shown to be independent of the reservoir, that is, $S$ is an inherent property of $A$ only. It is also shown that $S$ can be assigned absolute values that are non-negative, and that vanish for all the states encountered in mechanics. Moreover, and perhaps more importantly, by virtue of Eqs. 1 and 2, entropy remains invariant in any reversible adiabatic process of $A$, and increases in any irreversible adiabatic process of *A*. These conclusions are valid also for spontaneous processes, and for zero-net-effect interactions.

The dimensions of *S* depend on the dimensions of both energy and $c_R$. In due course we show that the dimensions of $c_R$ are independent of mechanical dimensions, and are the same as those of temperature [10].

Other rigorously proven theorems are: (i) In the course of interactions that change the state of a system *A* from $A_1$ to $A_2$, the difference $S_2 - S_1$ must equal the entropy exchanged with systems interacting with *A* plus a nonnegative amount generated spontaneously within *A*; the latter amount is called *entropy generated by irreversibility*; (ii) The minimum value of entropy is zero; (iii) If a system is in a stable equilibrium state, then and only then the entropy is an analytic function of the form $S(E, \bm{n}, V)$, and the concepts of temperature *T*, total potentials $\mu_i$ for i = 1, 2, …r, and pressure *p* are defined in terms of partial derivatives of $S(E, \bm{n}, V)$; (iv) For states that are not stable equilibrium, *T*, $\mu_i$, and *p* are undefinable and meaningless; (v) *Work* is an interaction that involves only the exchange of energy between the system and other systems in its environment; (vi) *Heat* is an interaction that involves only the exchange of energy and entropy between either a system and one or more reservoirs, and/or between two systems behaving as black body radiators, and differing infinitesimally in temperature; (vii) Neither work nor heat are contained in a system; (viii) Any expression that purports to represent entropy must conform to eight conditions or equivalently have the following characteristics:

(1) The expression must be well defined for every system (large or small), and every state (stable equilibrium or not stable equilibrium).
(2) The expression must be invariant in all reversible adiabatic processes, and increase in any irreversible adiabatic process.
(3) The expression must be additive for all systems and all states.
(4) The expression must be non-negative, and vanish for all the states encountered in mechanics.
(5) For given values of energy, amounts of constituents, and parameters, one and only one state must correspond to the largest value of the expression.
(6) For given values of the amounts of constituents and parameters, the graph of entropy versus energy of stable equilibrium states must be concave and smooth.
(7) For a composite *C* of two subsystems *A* and *B*, the expression must be such that the entropy maximization procedure for *C* [criterion no. (5)] yields identical thermodynamic potentials (for example, temperature, chemical potentials, and pressure) for all three systems *A*, *B*, and *C*.
(8) For stable equilibrium states, the expression must reduce to relations that have been established experimentally and that express the entropy in terms of the values of energy, amounts of constituents, and parameters, such as the relations for ideal gases.

It is noteworthy that, except for criteria (1) and (4), we can establish the remaining six criteria by reviewing the behavior of the entropy of classical thermodynamics.

The definition of entropy introduced here differs radically from and is more general than the entropy presented in practically all textbooks on physics and thermodynamics. Despite these differences, for thermodynamic equilibrium states, it has the same values as those listed in existing tables.

In the new exposition, the third law avers that: *For each given set of values of* **n** *and V of system A (without a finite upper limit on energy) there exists one stable equilibrium state with* *zero temperature, or infinite inverse temperature. For a system with both a lower and an upper limit on energy, such as a spin system, there exist two stable equilibrium states with zero temperatures, or equivalently* $-\infty \leq 1/T \leq \infty$.

Neither the statements of the three laws nor the proofs of any of their theorems require any considerations about numerical difficulties that prevent us from making explicit calculations, and about statistical measures of ignorance (or lack of information), or any restrictions to systems of specific sizes and specific numbers of degrees of freedom, or any limitations to states of specific types. So a statistical interpretation of thermodynamics is unwarranted, and a restriction to specific states unjustifiable.

## II.2   An energy versus entropy graph

At an instant in time, a state can be represented by a point in a multidimensional space with one axis for each amount of constituent, volume, and each independent property. Such a representation, however, is unwieldy because the number of independent properties of any system, even a system consisting of one particle only, is infinite. Nevertheless, useful information can be captured by first cutting the multidimensional state space by a hypersurface corresponding to given values of each amount of constituent and the volume, and then projecting the cut on an energy versus entropy plane. For system $A$ without upper bound on energy, it is proven that the projection must have the shape of the cross-hatched area in Figure 1.

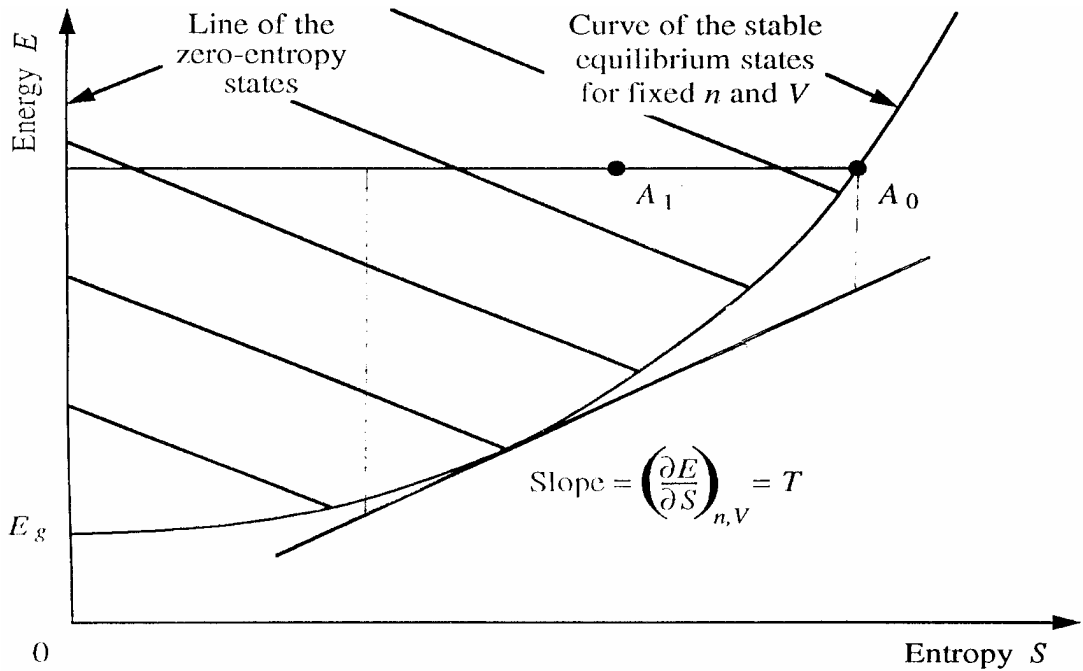

Figure 1. Energy versus entropy graph.

A point either inside the cross-hatched area or on the line $S = 0$ represents the projections of an infinite number of states. Each such state has the same values of amounts of constituents $n$, volume $V$, energy $E$, and entropy $S$ but differing values of other properties, and is not a stable equilibrium state. In particular, the line (and more generally the surface) $S = 0$ represents all the states regularized by the "known laws" of physics. The convex curve represents classical thermodynamics for given $n$ and $V$. Each point on the curve corresponds to one and only one stable equilibrium state. For any such state, the value of any property is determined solely by the values of $E$, $n$, and $V$. Many theorems of the laws of thermodynamics can be elegantly and simply illustrated on the $E$ versus $S$ diagram [10]. Projections of other cuts of the multidimensional state space on other planes, such as $E$ versus $V$, or $E$ versus the amount of a constituent, are possible. Each results in a graph that provides visual illustrations of different aspects of the new exposition.

### II.3   A thermodynamic exorcism of Maxwell's demon

Maxwell is one of the great scientists who believed that all physical phenomena are mechanical, but numerical difficulties with macroscopic systems force us to abandon the mechanical explanation and resort to the statistical method. He said: "One of the best established facts in thermodynamics is that it is impossible in a system enclosed in an envelope which permits neither change of volume nor passage of heat, and in which both the temperature and the pressure are everywhere the same, to produce any inequality of temperature or of pressure without the expenditure of work. Now let us suppose that such a vessel is divided into two portions $B$ and $C$ by a division in which there is a small hole, and that being who can see the individual molecules, opens and closes this hole, so as to allow only the swifter molecules to pass from $B$ to $C$, and only the slower ones to pass from $C$ to $B$. He will thus, without expenditure of work, raise the temperature of $C$ and lower that of $B$, in contradiction to the second law of thermodynamics." This being was later named a demon by Thomson.

Hundreds of papers and several books have been written over the past century, all claiming to prove that the demon cannot violate the second law. In our view, none of these publications has proven what is claimed because none addresses the problem posed by Maxwell. In each publication, either the demon or the environment of the vessel, or both experience some effects in sharp contrast to Maxwell's specification that such effects are not needed by an omniscient and omnipotent demon that accomplishes his task without expenditure of work, and, therefore, without any contribution whatsoever. One may think that such a specification is too restrictive and unrealistic. Nevertheless, this is Maxwell's conception.

In the new exposition, the exorcism satisfies Maxwell's specification, is definitive, and applies even if the molecules do not behave as a perfect gas and, therefore, cannot be treated individually [11]. The proof of these assertions can be readily illustrated by means of the $E$ versus $S$ diagram for the air molecules. Starting from stable equilibrium state $A_0$, the demon is asked to sort the air molecules into swift and slow without any changes in the values of the energy, the amount of air and the volume, and without any change either of his state or, more generally, of the state of the environment. If this were possible, the final state of $A$ would be $A_1$ (Fig. 1), that is a state with the same values of $E$, $n$, and $V$ as those of $A_0$, but less entropy than that of $A_0$. But we have proven that entropy is a nondestructible, nonstatistical property of the molecules of $A$. Accordingly, the demon cannot reduce the entropy without compensation. It is clear that this impossibility has nothing to do with either entropy generated by irreversibility, shortcomings of the demon's procedures and equipment, or collection and discard of information.

Equivalently, if the demon is regarded either as a cyclic machine or a perpetual motion machine of the second kind (PMM2), then his ultimate task is to extract only energy from system $A$ and, thus, change state $A_0$ to a state of smaller energy and equal or larger entropy than those of $A_0$. But under the specified conditions, the graph in Figure 1 shows that there exists no such state.

Some authors claim that the demon is infeasible even if the initial state of $A$ is not stable equilibrium [12]. This claim is also erroneous. If the initial state $A_1$ is not stable equilibrium and, therefore, lies somewhere within the cross-hatched area in Figure 1, then even an incompetent demon could extract only energy from $A$ without violating the laws of thermodynamics. Among a myriad of examples that illustrate the remarks just cited, a simple one is the work done by the small battery encapsulated in your or my wrist watch!

**II.4    Reversibility and the age of the universe**

Some scientists believe that we can expect to see unusual events such as gases unmixing themselves, only if we wait for times inconceivably long. There are lots of experiments that contradict this belief. For example, a well insulated bucket initially containing hot and cold water. Upon interacting only with each other, the hot and cold water become lukewarm and of course the process is irreversible. However, we can always restore the hot and cold parts over a very short period of time by means of cyclic machinery which leaves the energy of the environment intact but increases its entropy even if the processes are thermodynamically perfect, i.e., reversible. Moreover, the restoration of the initial state of the water is independent of the speed at which it is achieved, and involves neither velocity reversals nor any special information.

Another example is a high quality charged battery wrapped in excellent insulation and left idle on a shelf. After a few years, the battery is found to be dead because of internal discharge at constant energy. At that time, we can restore the initial state of the battery over a period of time much shorter than the time required for the completion of the spontaneous internal discharge. The spontaneous discharge is irreversible. Upon completing the recharging process, the energy of the environment is unchanged but its entropy increases even if the recharging is perfect, and occurs over a short or long period of time.

**III.    A NEW EXPOSITION OF QUANTUM THERMODYNAMICS**

**III.1    Foundations**

In this section we present a brief summary of a nonrelativistic quantum theory that differs from the presentations in practically every textbook on the subject. The key differences are the discoveries that for a broad class of quantum-mechanical problems: (i) The probabilities associated with ensembles of measurement results at an instant in time require a mathematical representation delimited by but more general than a wave function or projector; and (ii) The evolution in time of the new mathematical representation requires a nonlinear equation of motion delimited by but more general than the Schrödinger equation.

In response to the first difference, Hatsopoulos and Gyftopoulos [13] observed that there exist two classes of quantum problems. In the first class, the probabilities associated with measurement results are fully described by a wave function or projector, whereas in the second class the probabilities require a density operator $\rho$ that involves no statistical averaging over

projectors – no mixtures of quantum and statistical probabilities. The same result emerges from the excellent review of the foundations of quantum mechanics by Jauch [14]. In addition, the recognition of this difference eliminates the "*monstrosity*" of the concept of mixed state that concerned Schrödinger [15] and Park [16], and provides the link between quantum theory and thermodynamics without resort to statistics. This link extends the realm of quantum theory to states encountered in thermodynamics, and thermodynamic principles to quantum phenomena.

In either the case of a projector $\rho^2 = \rho$, or of a nonstatistical density operator $\rho^2 < \rho$, the pictorial representation of ρ is a homogeneous ensemble, that is, each member of the ensemble is characterized by the same ρ as the ρ of the whole ensemble. This fundamental difference must be contrasted to heterogeneous ensembles where ρ is a statistical average of projectors $\rho_i \neq \rho$.

For unitary evolutions of ρ in time, Hatsopoulos and Gyftopoulos [13] postulate that ρ obeys the equation

$$\frac{d\rho}{dt} = -\frac{i}{h}[H,\rho] \qquad (4)$$

for both isolated systems (H independent of time) and nonisolated systems (H dependent on time).

It is noteworthy that though Eq. 4 looks like the von Neumann equation of statistical quantum mechanics, here it must be postulated because ρ is not a statistical mixture of projectors and, therefore, Eq. 4 cannot be derived as a statistical average of Schrödinger equations.

As it is well known, the processes described by Eq. 4 are reversible adiabatic. If there exist constants of the motions of all the reversible adiabatic processes described by Eq. 4, each such constant must be a functional solely of the eigenvalues of ρ because these are the only quantities that remain invariant in the course of all unitary transformations with respect to time.

Using the conclusion just cited, and the eight conditions discussed in Section II.1, Gyftopoulos and Çubukçu [17] prove that the only expression for entropy that is acceptable is the

$$S = -kTr\rho \ln \rho \qquad (5)$$

provided that ρ is purely quantum mechanical, and not a mixture of quantum mechanical and statistical probabilities as in the case of the von Neumann entropy.

**III.2   A quantum thermodynamic exorcism of Maxwell's demon**

A theorem of quantum thermodynamics is that each molecule of a system in a thermodynamic equilibrium state has zero value of momentum, that is, each molecule is at a standstill and, therefore, there are no molecules to be sorted as swift and slow. The proof of this assertion is given in [23]. It is noteworthy that each molecule is at a standstill even in a system that is in an equilibrium state that is not stable.

The idea that in an equilibrium state each molecule or atom is at a standstill has been the subject of many nonscientific criticisms. In Ref. [23] I provide fully documented and rational responses to all the criticisms that I am aware of.

### III.3 Pictorial illustration of entropy

In many textbooks [24, 25], the probability density function $(\rho = \rho^2)$ of the spatial coordinates is interpreted as the shape of the constituents of a system. Gyftopoulos observed [26-28] that the same interpretation of the spatial shape applies to the probabilities derived from density operators $\rho > \rho^2$.

It follows that the entropy of thermodynamics (Eq. 5) is a measure of the quantum-theoretic spatial shape of constituents. Examples of this interpretation and how entropy changes from zero to larger values as the spatial shape $\rho^2 \leq \rho$ changes are given in [26-28]. For example, for one particle confined in either a one-dimensional or a two-dimensional infinitely deep potential well, and having a fixed energy, the spatial shapes are oscillatory, and become flat as the particle reaches a thermodynamic equilibrium state. Similarly, an electron of a hydrogen atom begins with beautiful but complicated spatial shapes [25] and ends up with a perfect spherical shape if the electron is in a thermodynamic equilibrium state.

## IV. THE EQUATION OF MOTION OF QUANTUM THERMODYNAMICS

### IV.1 Introduction

In response to the second difference cited in III.1, Beretta in his doctoral dissertation [18, 19] conceived a nonlinear equation of motion for the nonstatistical density operator $\rho$. The equation consists of a linear part that tends to drive the operator $\rho$ along a unitary isoentropic evolution and maintains constant each eigenvalue of $\rho$, and a conservative but dissipative force that pulls $\rho$ toward the path of steepest entropy ascent. In what follows, we discuss the simplest application of the equation. More information and applications are given in [20-22].

### IV.2 One particle approximation for a Boltzmann gas

As an illustration of the Beretta equation, we consider an isolated system composed of non-interacting identical particles with single-particle energy eigenvalues $e_i$ for i = 1, 2, ..., N where N is finite and the $e_i$'s are repeated in case of degeneracy. As done by Beretta [22], we restrict our analysis on the class of dilute-Boltzmann-gas states in which the particles are independently distributed among the N (possibly degenerate) one-particle energy eigenstates. In density operator language, this is tantamount to restricting the analysis on the subset of one-particle density operators that are diagonal in the representation in which also the one-particle Hamiltonian operator is diagonal. We denote by $p_i$ the probability of the i-th energy eigenstate, so that the per-particle energy and entropy functionals are given by the relations

$$E = \sum_{i=1}^{N} e_i p_i \qquad S = -k \sum_{i=1}^{N} p_i \ln p_i \qquad \sum_{i=1}^{N} p_i = 1 \qquad (6)$$

As in all paradigms of physics, the nonlinear equation of motion maintains the initially zero probabilities equal to zero, whereas the rates of change of the nonzero probabilities are given by

$$\frac{dp_j}{dt} = -\frac{1}{\tau} \frac{\begin{vmatrix} p_j \ln p_j & p_j & e_j p_j \\ \sum p_i \ln p_i & 1 & \sum e_i p_i \\ \sum e_i p_i \ln p_i & \sum e_i p_i & \sum e_i^2 p_i \end{vmatrix}}{\begin{vmatrix} 1 & \sum e_i p_i \\ \sum e_i p_i & \sum e_i^2 p_i \end{vmatrix}} \quad \text{for } i, j = 1, 2, \ldots, N \qquad (7)$$

where $\tau$ is a scalar time constant or functional.

The solutions of these equations are well-behaved in the sense that they satisfy both all the conditions given in [23], and have the following general features: (i) They conserve the energy and trace of $\rho$ ; (ii) They preserve the non-negativity of each $p_i$; (iii) They maintain the non-negative rate of entropy generation; (iv) They maintain the dimensionality of the density operator; (v) They drive any arbitrary initial density operator $\rho(0)$ toward the partially canonical equilibrium density operator $\rho(\infty)$ with time independent eigenvalues in the energy representation

$$p_j^{pe}(E, t=\infty) = \frac{\exp(-\beta^{pe} e_j)}{\sum_{i=1}^{N} \exp(-\beta^{pe} e_i)} \qquad (8)$$

where the value of $\beta^{pe}$ is determined by the initial condition $\sum_{i=1}^{N} e_i p_i^{pe}(E) = E = E(\rho(0))$, and the superscript pe is used to indicate that the system is in an unstable or, so-called, partial equilibrium state.

Among all the equilibrium states just cited there exists one and only one that is stable (se) and corresponds to the largest value of the entropy for the given value of energy $E$, and for which the eigenvalues of the density operator in the energy representation are given by the relations – canonical distribution

$$p_j^{se}(E) = \frac{\exp(-e_j/kT(E))}{\sum_{i=1}^{N} \exp(-e_i/kT(E))} \qquad (9)$$

where $T(E)$ is shown to be equal to the derivative of energy with respect to entropy of stable equilibrium states of the Boltzmann gas at energy $E$. By definition the derivative just cited is called temperature.

For a general nonequilibrium state, the rate of entropy generation may be written as a ratio of Gram determinants in the form

$$\frac{dS}{dt} = -\frac{k}{\tau} \frac{\begin{vmatrix} \sum p_i (\ln p_i)^2 & \sum p_i \ln p_i & \sum e_i p_i \ln p_i \\ \sum p_i \ln p_i & 1 & \sum e_i p_i \\ \sum e_i p_i \ln p_i & \sum e_i p_i & \sum e_i^2 p_i \end{vmatrix}}{\begin{vmatrix} 1 & \sum e_i p_i \\ \sum e_i p_i & \sum e_i^2 p_i \end{vmatrix}} \geq 0 \qquad (10)$$

where the non-negativity follows from the well-known properties of Gram determinants.

Given any initial density operator, it is possible to solve the equation of motion for all values of time, that is $-\infty < t < \infty$. In the limit $t \to \infty$ the trajectory approaches a largest entropy equilibrium state with a density operator that is canonical over the energy eigenstates initially included in the analysis. An exception to this conclusion is the case of the initial density operator being a projector $\rho = \rho^2$. Then the evolution in time follows the Schrödinger equation, and is unitary and reversible, except if the projector is an energy eigenprojector which is stationary.

As stated earlier, in the unified quantum theory of mechanics and thermodynamics without statistical probabilities, the three laws of thermodynamics introduced in Section II need not be introduced explicitly because they are theorems of the new exposition of quantum thermodynamics. This fact is analogous to the derivation of momentum and kinetic energy conservations as theorems of Newton's equation of motion of classical mechanics.

### IV.3    Discussion of views about thermodynamics

Some views of preeminent scientists about the nature of thermodynamics are reviewed in the light of the two novel expositions of the subject in Ref. [8]. In particular, comments made by Boltzmann [29], Brillouin [30], Feynman [31], Penrose [32], Denbigh [33], and Lebowitz [34] are reviewed and found to misrepresent the principles and theorems of both the exposition of thermodynamics without quantum considerations, and/or the quantum theory of mechanics and thermodynamics without statistical probabilities.

### V.    CONCLUSIONS

The most important conclusion of this work is that entropy is a quantum-theoretic, inherent, nonstatistical property of any system (large or small, including a one spin system), in any state (thermodynamic equilibrium or not thermodynamic equilibrium). Another conclusion is that thermodynamics is a general quantum theory that enlarges the realm of quantum mechanics from zero entropy physics to physics for nonzero values of entropy, and the realm established for conditions of thermodynamic equilibrium. Thus, the conception of the unification of mechanical and thermodynamic concepts eliminates the need for the ideas of randomness, disorder, lack or erasure of information, and difficulty of performing complicated calculations.